\documentclass
[aps,twocolumn,showpacs,superscriptaddress,tightenlines]{revtex4}%
\usepackage{epsfig,rotating}
\usepackage{amsmath}
\usepackage{amsfonts}
\usepackage{amssymb}
\usepackage{graphicx}%
\providecommand{\U}[1]{\protect\rule{.1in}{.1in}}
\providecommand{\U}[1]{\protect\rule{.1in}{.1in}}
\begin{document}
\title{Blurred femtoscopy in two-proton decay}
\author{C.\thinspace A.~Bertulani}
\email{carlos_bertulani@tamu-commerce.edu}
\affiliation{Department of Physics, Texas A\&M University, Commerce, TX 75429, USA}
\author{M.S. Hussein}
\email{mhussein@mpipks-dresden.mpg.de} \affiliation{Instituto de
F\'{\i}sica, Universidade de S\~{a}o Paulo, C.P. 66318, 05389-970
S\~{a}o Paulo, Brazil} \affiliation{and Max-Planck-Institut f\"ur
Physik komplexer Systeme, 01187 Dresden, Germany}
\author{G. Verde}
\email{verde@ct.infn.it} \affiliation{INFN, Sezione di Catania Via
S. Sofia 64 (Cittadella Universitaria) I-95123 Catania, Italy}
\date{\today }

\begin{abstract}
We study the effects of final state interactions in two-proton
emission by nuclei. Our approach is based on the solution the
time-dependent Schr\"odinger equation. We show that the final
relative energy between the protons is substantially influenced by
the final state interactions. We also show that alternative
correlation functions can be constructed showing large sensitivity
to the spin of the diproton system.

\end{abstract}

\pacs{23.50.+z, 27.20.+n, 27.40.+z, 29.40.Cs}
\maketitle

%\narrowtext

\label{intr}

Two-proton emission has been observed for numerous excited states in
nuclei, populated both in $\beta$ decay and in nuclear reactions
\cite{Bl05,Gi07,Mu07,Mu06,Mi07}. Although these decays are thought to be
sequential one-proton emissions proceeding through states in the
intermediate nucleus \cite{Bai96,Fyn00}, there is an intriguing
possibility \cite{Gol60} that the diproton ($^{2}$He) correlation
may play an important role in the mechanism of the two-proton
emission. This has been nicely demonstrated in the analysis of the two-proton
decay of the $21^{+}$ isomeric state in $^{94}$Ag, where 19 events were clearly
assigned to the simultaneous emission of two correlated protons \cite{Mu06}.
The traditional idea of diproton radioactivity is due to
the pairing effect. Two protons form a quasiparticle (diproton)
under the Coulomb barrier and this facilitates penetration. In a
more formal description, one has a system with two valence protons
in the same shell and coupled to $J^{\pi}=0^{+}$. This question,
being still open, continues to motivate studies in this field. In
order to assess this information, it is necessary to understand
final state interactions between the protons and between each proton
and the daughter nucleus.

In nuclear decays the emission of correlated, identical, particles is
sensitive to the geometry of the system. Measurements of correlation functions
are often performed with charged particle pairs, which interact via the
short-range nuclear interaction and the long-range Coulomb interaction and
they also interact with the remaining source. As a result, theoretical
corrections are needed to subtract the final state interactions (FSI) before
one can extract any useful information about the emitting source from the
measurements \cite{Bow91,BBM96,Bar98,Sin98}. At first sight, the FSI can be
regarded as a contamination of \textquotedblleft pure\textquotedblright%
\ particle correlations. However, it should be noted that the FSI depend on
the structure of the emitting source and thus provide information about source
dynamics as well.

Two-proton decay in s-wave states can also be used for testing
quantum mechanics versus local realism by means of Bell's
inequalities \cite{Ber03}. Since the final state of the two protons
can be either in a singlet or in a triplet state, their wavefunction
is spin entangled. The identification of the spins of the proton in
two detectors separated far away would be useful to test the
Einstein--Podolski--Rosen (EPR) paradox \cite{EPR,Bell}. In fact,
these tests should be performed in different and complementary
branches of physics to avoid the loopholes encountered in photon
experiments. The advantage of using massive Fermions to test
Bell-type inequalities is that the particles are well localized and
the spin state of the pair can be well established by measuring the
internal energy of the two-proton system. However, the validity of
this method highly depends on our ability to treat FSI. Coincidence
measurements of the two proton momenta require knowledge of FSI in
order to extract information about their original wavefunction. Here
will propose a new method to calculate FSI based on the numerical
solution of the time-dependent Schr\"odinger equation. We hope with
that to get a quantitative estimate of the FSI and how they can be
used to address the points raised above. Though tests of Bell's
inequality using proton-proton spin correlation in low energy
scattering conformed to quantum mechanics (see, e.g., \cite{Lam76}),
one would still like to find other means to perform the verification
of the complete  nature of quantum mechanics.

One should distinguish this work from Hanbury-Brown-Twiss (HBT)
studies in high enery nucleus nucleus collisions. Indeed, in the
case of HBT, the whole game is played by FSI. FSI are the means by
which one can determine information about the source. In this case
FSI is not viewed as a "contamination". The important difference in
our case is that we are interested in a case where the protons are
not emitted chaotically like in the case of HIC (Heavy-Ion
Collisions). We are studying nuclear structure. In HIC (and HBT)
protons are assumed to be emitted independently and chaotically from
the source (any information about their initial spins is lost and
they are assumed to be "evaporated" with a boiling pot). Then, there
are no initial state correlations and FSI make the whole physics. In
the case of two-proton radioactivity, the emission of the two
protons is not chaotic because their correlation function keeps
memory of their spin admixture and wave function in the parent
nucleus.

We consider first a single proton described at the initial time by a localized
wave-packet $\psi_{0}(\mathbf{r}_{1})$. The probability amplitude to find the
proton at the detector with momentum $\mathbf{p}_{1}$ is given by%
\begin{equation}
\mathcal{A}\left(  \mathbf{p}_{1},\mathbf{r}_{1}\right)  =\int d\mathbf{r}
\chi^{(+)}(\mathbf{p}_{1},\mathbf{r})K\left(  \mathbf{r,r}_{1}\right)
\psi_{0}(\mathbf{r}_{1}), \label{apr}%
\end{equation}
where $\chi^{(+)}(\mathbf{p}_{1},\mathbf{r})$ is an asymptotic outgoing
Coulomb wave with energy $E=\mathbf{p}_{1}^{2}/2m_{p}$, and $K\left(
\mathbf{r,r}_{1}\right)  $ is the propagator which accounts for the time
evolution of the particle from the source to the detector.

We now look at the case of two-protons interacting with the residual nucleus
and between themselves. We will consider the distortion caused by the Coulomb
plus nuclear interaction between each proton $i$ with the nucleus,
$V_{C}(\mathbf{r}_{i})+V_{N}(\mathbf{r}_{i})$, and between themselves,
$v_{C}^{12}(\mathbf{r})+v_{N}^{12}(\mathbf{r)}$, where $r_{i}$ is the
coordinate of proton $i$, and $\mathbf{r}$ is their relative coordinate. The
proton-nucleus interaction, $V_{N}(\mathbf{r_{i})}$ yields smaller final state
interaction effects than the Coulomb counterpart.

We adopt a classical description of the center-of-mass motion for the
two-protons and solve the time-dependent Schr\"{o}dinger equation for the
relative motion between them. The Coulomb field that distorts the relative
motion of the particles is given by
\begin{equation}
V_{C}(t)=Ze^{2}\left(  \frac{1}{\left\vert \mathbf{r}_{1}-\mathbf{R}%
(t)\right\vert }-\frac{1}{\left\vert \mathbf{r}_{2}-\mathbf{R}(t)\right\vert
}-\frac{2}{R(t)}\right)  , \label{vct}%
\end{equation}
where $Z$ is the charge of the daughter nucleus and $\mathbf{r}_{1}$\ and
$\mathbf{r}_{2}$\ are the positions of the protons with respect to the center
of the nucleus (nuclear recoil is neglected). $V_{C}(t)$ acts on the relative
position $\mathbf{r=r}_{2}-\mathbf{r}_{1}$ through the transformations
$\mathbf{r}_{1}=\mathbf{R}-\mathbf{r}/2$\ and $\mathbf{r}_{2}=\mathbf{R}%
+\mathbf{r}/2$.

One can perform a multipole expansion of this interaction and for $r$ smaller
than $R(t)$ one can express the result in terms of a multipole-dependent
effective charge, $e_{L}=e\left[  \left(  -1/2\right)  ^{L}+\left(
1/2\right)  ^{L}\right]  \label{el}$ where $L$ is the multipole degree. The
dipole field ($L=1$) is only important for particles with different
charge-to-mass ratios while the quadrupole field is dominant when these ratios
are equal (e.g. for two-proton emission). For the quadrupole interaction,
$e_{L=2}=e/2$ and
\begin{equation}
V_{C}\left(  t\right)  =\frac{Ze^{2}}{2}\frac{r^{2}}{R^{3}(t)}P_{2}\left(
\cos\theta\right)  , \label{vct2}%
\end{equation}
where $\theta$ is the angle between \textbf{R} and \textbf{r} and $P_{2}$ s
the Legendre polynomial of order 2. We now assume that the protons are
produced simultaneously and nearly at rest at position $2a_{0}$ and time
$t=0$. Their center-of-mass follows a radial trajectory described by%
\begin{equation}
R(t)=\frac{a_{0}}{2}\left(  \cosh w+1\right)  ,\ \ \ \ \ \ \ t=\frac{a_{0}%
}{2v}\left(  \sinh w+w\right)  , \label{rtt}%
\end{equation}
where the asymptotic velocity is given by $v=\sqrt{E/m_{p}}$,  $E$
is the two-proton decay energy, and $a_{0}=e^{2}/2E$. This assumes
that the relative energy between the protons is much smaller than
$E$, which is not a good approximation, as we will show later. It is
important to notice that eqs. \ref{rtt} only account for the motion
of the protons after they emerge from inside the nucleus through the
Coulomb barrier and propagate from the closest distance $2a_{0}$ to
infinity. Hence, our calculations neglect what happens during the
tunneling process and treat only the external motion. Hence,
neglecting the proton-nucleus strong FSI is justified.

We can still use eq. \ref{apr} to calculate the probabilities for relative
motion of the protons, with the wavefunction for the relative motion given by
$\Psi(\mathbf{r})=K\left(  \mathbf{r,r}_{0}\right)  \psi_{0}(\mathbf{r}_{0})$.
In the time dependent description, at time $t$ this wave function can be
expanded in spherical harmonics%
\begin{equation}
\Psi\left(  \mathbf{r}\right)  =\frac{1}{r}\sum_{lm}u_{lm}\left(  r,t\right)
Y_{lm}\left(  \widehat{\mathbf{r}}\right)  , \label{psir}%
\end{equation}
and the Schr\"{o}dinger equation, describing the time evolution of the
relative motion between the protons can be solved by the finite difference
method, calculating the wavefunction at time $t+\Delta t$ in terms of the
wavefunction at time $t$, according to the algorithm%
\begin{align}
u_{lm}\left(  t+\Delta t\right)   &  =\left[  \frac{1}{i\tau}-\Delta
^{(2)}+\frac{\Delta t}{2\hbar\tau}U\right]  ^{-1}\nonumber\\
&  \times\left[  \frac{1}{i\tau}+\Delta^{(2)}+\frac{\Delta t}{2\hbar\tau
}U+\frac{\Delta t}{\hbar\tau}S_{l^{\prime}m^{\prime};lm}\right]  u_{lm}\left(
t\right)  , \label{ulmt}%
\end{align}
where $\tau=\hbar\Delta t/m_{p}\left(  \Delta r\right)  ^{2}$. The second
difference operator is defined as
\begin{equation}
\Delta^{(2)}u_{lm}^{(j)}(t)=u_{lm}^{(j+1)}(t)+u_{lm}^{(j-1)}(t)-2u_{lm}%
^{(j)}(t), \label{dulmt}%
\end{equation}
with $u_{lm}^{(j)}(t)=u_{lm}(r_{j},t)$, where $r_{j}$ is a position in the
radial lattice. In eq. \ref{ulmt},\ $U=v_{C}^{12}\left(  r_{j}\right)
+v_{N}^{12}\left(  r_{j}\right)  $ is the Coulomb+nuclear interaction between
the two protons as a function of their distance, $r_{j}$, and the function
$S_{l^{\prime}m^{\prime};lm}$ is given by%
\begin{equation}
S_{l^{\prime}m^{\prime};lm}(r,t)=\sum_{l^{\prime}m^{\prime}}\left\langle
Y_{l^{\prime}m^{\prime}}\left\vert V_{C}(r,t)\right\vert Y_{lm}\right\rangle
u_{l^{\prime}m^{\prime}}\left(  r,t\right)  . \label{slm}%
\end{equation}

This method of solving the time-dependent equation is the same as used in ref.
\cite{BB94} for studying reacceleration effects in breakup reactions in
nucleus-nucleus collisions at intermediate energies. A grid adequate for our
purposes has 5000 spatial mesh points separated by 0.1 fm and 2000 time mesh
points separated by 0.5 fm/c.

We use the quantization-axis along the $\mathbf{R}(t)$ center-of-mass radial
trajectory. As a consequence, $P_{2}\left(  \cos\theta\right)  =\sqrt{4\pi
/5}Y_{20}\left(  \theta,\phi\right)  $, and one only needs to consider the
$m=0$ component of the spherical harmonics implicitly contained in the
potential $V_{C}(t)$. The initial $l=0$ state cannot develop a final $l=1$
component, and only $l=0$ (s-waves) and $l=2$ (d-waves) will be present in the
final state. Higher $l$ values will be small and need not be considered.

The proton-proton potential is taken as $v_{N}^{12}\left(  r\right)
+v_{C}^{12}\left(  r\right)  =e^{2}/r+v_{0}(b/r)\exp\left(  -r/b\right)  $.
The set of parameters $v_{0}=-46.124$ MeV and $b=1.1809$ fm yields the
proton-proton scattering length, $a_{p}=-7.8196$ fm and the effective range
$\rho_{0}=2.790$ fm, in accordance with experimental data. But we choose a
higher absolute value of $v_{0}$ which allows the presence of a single weakly
bound s-wave state. We use this localized wavefunction for the relative motion
of the two protons in the initial state: $u_{0}\equiv u_{l=0}\left(
r,t=0\right)  $. This is an artifact of the numerical method chosen as to
allow for a localization of the initial wavefunction. The observables
associated with the final state will depend on the binding energy, reflecting
the dependence on the initial average separation between them. The average
initial separation, $r_{0}$, and the binding energy, $B$, are approximately
related by $r_{0}=\hbar\left(  4Bm_{p}\right)  ^{-1/2}$.

As time evolves the initial state will acquire components in the continuum due
to the action of the interaction \ $V_{C}(t).$\ The continuum component
propagates as a wavepacket which moves away from the source with a final
asymptotic momentum $\mathbf{p}$. The continuum wavefunction is obtained by
removing the bound-state part from the solution of eq. \ref{ulmt}%
\[
\Psi_{c}(t)=\mathcal{N}\left[  \Psi(t)-\left\langle \Psi(t)|\Psi
_{0}\right\rangle \Psi_{0}\right]  ,
\]
where $\mathcal{N}$ normalizes the continuum wavefunction, $\Psi_{c}$, to unity.

The probability amplitude to find the protons with a final relative momentum
$\mathbf{p}$ is given by%
\begin{align}
\mathcal{A}\left(  \mathbf{p}\right)   &  =\left\langle \chi^{(+)}%
(\mathbf{p},\mathbf{r})|\Psi\left(  \mathbf{r},t\longrightarrow\infty\right)
\right\rangle \nonumber\\
&  =%
%TCIMACRO{\dint }%
%BeginExpansion
{\displaystyle\int}
%EndExpansion
u_{l=0}(r,t\longrightarrow\infty)H_{0}(pr)dr\nonumber\\
&  +%
%TCIMACRO{\dint }%
%BeginExpansion
{\displaystyle\int}
%EndExpansion
u_{l=2}(r,t\longrightarrow\infty)H_{2}(pr)dr,\label{pp}%
\end{align}
where
\begin{equation}
H_{l}(pr)=\exp\left[  i\left(  pr+\frac{l\pi}{2}-\eta\ln\left(  2pr\right)
+\sigma_{l}\right)  \right]  \label{chirp}%
\end{equation}
is the asymptotic Coulomb wavefunction for angular momentum $l$, with
$p\mathbf{=}\hbar k$, $\eta=e^{2}/\hbar v_{r}$, $v_{r}=\sqrt{4E_{r}/m_{p}}$ is
the asymptotic relative velocity, and $E_{r}=\hbar^{2}k^{2}/m_{p}$ the
relative energy.%

\begin{figure}
[ptb]
\begin{center}
\includegraphics[
height=1.9501in,
width=2.8452in
]%
{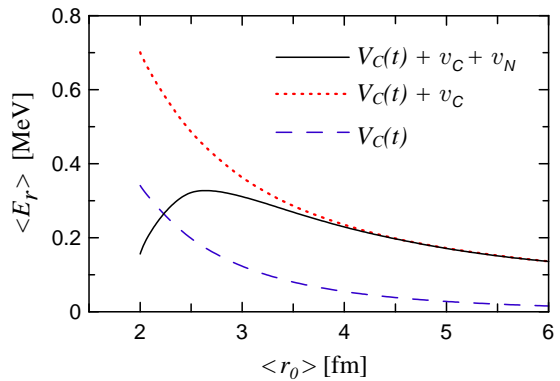}%
\caption{Average relative motion energy of the two protons as a function of
the average initial distance between them. The dashed curve is the final
proton-proton relative energy if their interaction is neglected. The dotted
curve includes the Coulomb repulsion between them and the solid curve includes
their Coulomb and nuclear interaction.}%
\label{fig1}%
\end{center}
\end{figure}

We consider two-proton decay from $^{45}$Fe with a decay energy of 1.1 MeV. In
figure \ref{fig1} we show the results for the average relative motion energy
$\left\langle E_{r}\right\rangle =\left\langle \mathcal{A}\left\vert
p^{2}/m_{p}\right\vert \mathcal{A}\right\rangle /\left\langle \mathcal{A}%
|\mathcal{A}\right\rangle $ as a function of the average initial distance
between the two protons. The dashed curve is the final proton relative energy
if their mutual interaction is neglected. The dotted curve includes the
Coulomb repulsion between the protons and the solid curve includes both their
Coulomb and nuclear interaction.

The physical reasons for the results shown in figure \ref{fig1} are
transparent: two strongly interacting particles emitted from the volume of
$r_{0}\approx2$ fm, when their \textquotedblleft own mean radius", reflected
by their attractive strong potential $v_{N}$, is about $b\approx1$ fm, should
sufficiently \textquotedblleft feel\textquotedblright\ each other through
mutual attraction. On the other hand, when the emission zone is much larger
than the particle mean radius, for instance if the radius is $r_{0}\approx6$
fm, the contribution of the short-range attraction is negligible and only the
long-range Coulomb repulsion acts, as is seen in fig. \ref{fig1}. It is also
clear that the final relative energy increases when the distance $r_{0}$
decreases. But as $r_{0}$ increases the contribution of the \textquotedblleft
tidal\textquotedblright\ Coulomb interaction of the diproton with the daughter
nucleus decreases much faster than the contribution of their own mutual
Coulomb repulsion.

It is also important to notice that all the contributions to the FSI
energies are not small compared to the decay energy $E$. This is
contrary to our initial assumption used to justify our dynamical
model. Hence, the results point to the important conclusion that
final-state interactions are very important in determining the
relation between the proton energies and the spatial distribution of
the protons in the decay process. The FSI contributions due to the
Coulomb tidal interaction depend on the square of the charge of the
daughter nucleus while the contribution of the strong force between
the protons is approximately independent of the nuclear mass.  Thus,
for lighter nuclei (e.g. $^{18}$Ne), the dashed curve in \ figure
\ref{fig1} becomes negligible.

Further aspects of the spatial configuration of the protons can be obtained in
light of the usual discussion in terms of two-particle interferometry, or
correlation functions. The relation between the center-of-mass coordinates and
the laboratory are given by%
\begin{align*}
\mathbf{R} &  =\frac{\left(  \mathbf{r}_{1}+\mathbf{r}_{2}\right)  }%
{2},\ \ \ \ \ \mathbf{r}=\mathbf{r}_{1}-\mathbf{r}_{2},\\
\mathbf{P} &  =\mathbf{p}_{1}+\mathbf{p}_{2},\ \ \ \ \ \mathbf{p}%
=\frac{\left(  \mathbf{p}_{1}-\mathbf{p}_{2}\right)  }{2}.
\end{align*}
\newline The probability amplitude to find one of the protons with momentum
$\mathbf{p}_{1}$ is given by $\mathcal{A}_{1}\left(  \mathbf{p}_{1}%
,\mathbf{r}_{1}\right)  =\mathcal{A}\left(  \mathbf{p}_{1}-\mathbf{P}%
/2,\mathbf{r}_{1}\right)  $. The center of mass momentum, $\mathbf{P}%
$\textbf{,} is set by the decaying energy and the assumption that it follows
an outgoing radial motion, i.e. $\mathbf{P}=\hat{\mathbf{R}}\sqrt{E/m_{p}}$,
where $\hat{\mathbf{R}}$ is the unit vector along the radial direction.

Next we show that one can disentangle the contributions of singlet and triplet
spin final states of the two-proton system by measuring momentum correlations.
This will prove to be a useful method since a direct measurement of the spin
orientations of each proton is by far more complicated. The application of the
method is very general as it only relies on measured quantities, independent
of the models for the treatment of FSI.

The protons are identical particles and their detection requires the
consideration of their quantum statistical properties. If proton 1 is detected
with momentum $\mathbf{p}_{1}$ and proton 2 is detected with momentum
$\mathbf{p}_{2}$, the probability amplitude for this is given by product
$\mathcal{A}_{1}\left(  \mathbf{p}_{1},\mathbf{r}_{1}\right)  \mathcal{A}%
_{2}\left(  \mathbf{p}_{2},\mathbf{r}_{2}\right)  $. Because of the
indistinguishability of the particles, the probability amplitude must be
symmetric with respect to the interchange of two particles if they are in a
spin-singlet state ($S=0$), and antisymmetric if they are in a spin-triplet
state ($S=1$). The normalized probability amplitude becomes%
\begin{align*}
&  \Lambda^{\left(  \pm\right)  }\left(  \mathbf{p}_{1},\mathbf{p}%
_{2},\mathbf{r}_{1},\mathbf{r}_{2}\right)  =\frac{1}{\sqrt{2}}\left[
\mathcal{A}_{1}\left(  \mathbf{p}_{1},\mathbf{r}_{1}\right)  \mathcal{A}%
_{2}\left(  \mathbf{p}_{2},\mathbf{r}_{2}\right)  \right. \\
&  \left.  \pm\mathcal{A}_{1}\left(  \mathbf{p}_{2},\mathbf{r}_{1}\right)
\mathcal{A}_{2}\left(  \mathbf{p}_{1},\mathbf{r}_{2}\right)  \right]  ,
\end{align*}
where the plus or minus sign is the spin-singlet and spin-triplet state, respectively.

The two-particle momentum distribution $P(\mathbf{p}_{1},\mathbf{p}_{2})$ is
the probability to measure a nucleon having momentum $\mathbf{p}_{1}$ in
coincidence with the measurement of the other nucleon having momentum
$\mathbf{p}_{2}$. It is defined as%
\begin{align}
P(\mathbf{p}_{1},\mathbf{p}_{2})  &  =%
%TCIMACRO{\dint }%
%BeginExpansion
{\displaystyle\int}
%EndExpansion
d^{3}r_{1}d^{3}r_{2}\Big\vert\left\vert \Lambda^{\left(  +\right)  }\left(
\mathbf{p}_{1},\mathbf{p}_{2},\mathbf{r}_{1},\mathbf{r}_{2}\right)
\right\vert ^{2}\nonumber\\
&  \pm\mathcal{M}\Lambda^{\left(  -\right)  }\left(  \mathbf{p}_{1}%
,\mathbf{p}_{2},\mathbf{r}_{1},\mathbf{r}_{2}\right)  ^{2}\Big\vert^{2},
\end{align}
where $\mathcal{M}$ is the mixing parameter, determining the relative
contribution of the triplet state. The correlation function $C(\mathbf{p}%
_{1},\mathbf{p}_{2})$ is defined as the ratio of the probability for the
coincidence of $\mathbf{p}_{1}$ and $\mathbf{p}_{2}$ relative to the
probability of observing $\mathbf{p}_{1}$ and $\mathbf{p}_{2}$ separately,%
\[
C(\mathbf{p}_{1},\mathbf{p}_{2})=\frac{P(\mathbf{p}_{1},\mathbf{p}_{2})}%
{P_{1}\left(  \mathbf{p}_{1}\right)  P_{2}\left(  \mathbf{p}_{2}\right)  }.
\]

Let us assume for the moment the that protons suddenly emerge from
the nucleus and that their intrinsic wavefunction is in a pure
entangled state. If their
wavefunction is approximated by $\exp(i\mathbf{p}_{1}.\mathbf{r}_{1}%
)\exp(i\mathbf{p}_{2}.\mathbf{r}_{2})\pm\exp(i\mathbf{p}_{2}.\mathbf{r}%
_{1})\exp(i\mathbf{p}_{1}.\mathbf{r}_{2})$ it will lead to destructive or
constructive interferences. If we also assume a gaussian source of size
$r_{0}\equiv\sqrt{\left\langle r^{2}\right\rangle }$, the correlation function
without final state interactions would be given by%
\begin{equation}
C(\mathbf{p}_{1},\mathbf{p}_{2})   \equiv C\left(  q\right)
=f\left(  \Delta p\right)  \left[  1\pm\exp\left(  -\Delta
p^{2}r_{0}^{2}/\hbar^{2}\right)
\right]   ,\label{cfree}%
\end{equation}
where $\Delta p=p=|\mathbf{p}_{1}-\mathbf{p}_{2}|/2$. All other
features of the reaction mechanism are included in the function
$f\left(  \Delta p\right)  $. One can approximately account for the
Coulomb interaction between the protons
by using a Gamow function for $f\left(  \Delta p\right)  $:%
\begin{equation}
f\left(  \Delta p\right)  =\frac{2\pi\eta}{\exp\left(  2\pi\eta\right)
-1},\label{fgamow}%
\end{equation}
but no such simple estimate exists for the effect of the nuclear interaction.

According to eq. \ref{cfree}, for $\Delta pr_{0}/\hbar\ll1$ one should be able
to see a destructive interference for triplet final states and constructive
interference for singlet final states. It is thus appropriate to redefine the
correlation function in terms of the relative momentum between the protons, so
that the correlated ($\mathcal{C}$) and uncorrelated ($\mathcal{U}$)
measurements of protons 1 and 2 are defined by
\begin{align}
\mathcal{C}(\Delta p) &  =\int P\left(  \mathbf{p}_{1};\mathbf{p}_{1}%
+2\Delta\mathbf{p}\right)  d\mathbf{p}_{1}d\Omega_{p}%
\ ,\ \ \ \ \ \ \ \nonumber\\
\ \mathcal{U}(\Delta p) &  ={\frac{1}{N}}\int P\left(
\mathbf{p}_{1}\right) P\left(
\mathbf{p}_{1}+2\Delta\mathbf{p}\right)  d\mathbf{p}_{1}d\Omega
_{p}\ .\label{CU}%
\end{align}
The integration in $\Omega_{p}$ is over all orientations of $\Delta\mathbf{p}%
$. $P\left(  \mathbf{p}\right)  =\int P\left(  \mathbf{p};\mathbf{p}^{\prime
}\right)  d\mathbf{p}^{\prime}$ is the probability to measure the momentum
$\mathbf{p}$ for one of the protons, irrespective of what the momentum of the
other proton is. $N$ is the total number of particles measured, i.e., $N=\int
P\left(  \mathbf{p}\right)  d\mathbf{p}$.%

\begin{figure}
[ptb]
\begin{center}
\includegraphics[
height=1.9182in,
width=2.6671in
]%
{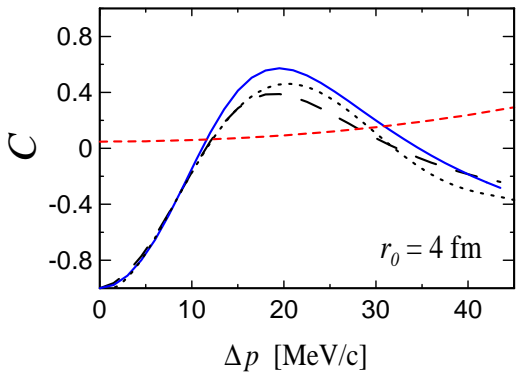}%
\caption{Correlation function, eq. \ref{R}, for the $^{45}$Fe decay
with $E=1.1$ MeV, as a function ot the relative momentum of the
protons.. The short-dashed curve is for the singlet state and all
others are for the triplet state. The dotted curve does not include
the proton-proton interaction, the long-dashed curve includes the
Coulomb interaction between the protons and the solid curve includes
both nuclear and Coulomb interaction between the
protons.}%
\label{fig2}%
\end{center}
\end{figure}

The new correlation function is defined as
\begin{equation}
C\left(  \Delta p\right)  ={\frac{\mathcal{C}(\Delta p)}{\mathcal{U}(\Delta
p)}}-1. \label{R}%
\end{equation}

In figure \ref{fig2} we show the correlation function for $^{45}$Fe
decay with $E=1.1$ MeV, as a function of the relative momentum of
the protons. The short-dashed curve is for the singlet state and all
others for the triplet state. The dotted curve does not include the
proton-proton interaction, the long-dashed curve includes the
Coulomb interaction between them and the solid curve includes both
nuclear and Coulomb interaction between the protons.

One sees that the properties of the correlation functions in the
singlet and triplet states are completely different. When $\Delta p$
is small the correlation function is negative only for the triplet
state. It is -1 at $\Delta p=0$ for the triplet state, whereas it is
close to zero for the singlet state. While for the former case the
correlation function crosses zero at two points, it does not have a
null point for the singlet case. \ It is also worthwhile mentioning
that the effect of the Coulomb interaction between the protons can
be switched off and the resulting correlation function
$\mathcal{C}(\Delta p)$ multiplied by the Gamow factor in eq
\ref{fgamow} yields a result (not shown in fig. \ref{fig2}) slightly
different than the long-dashed curve. In fact, the Gamow factor of
eq. \ref{fgamow} tends to
underestimate the Coulomb final state interaction between the protons.%

The method described above is directly applicable to determine the spin mixing
of final states in low-energy two-proton nuclear decay for $0^{+}%
\longrightarrow0^{+}$ transitions. In this case the final spin wave function
of the pair equals that of the initial wave function. In particular, when
singlet states are identified, spin-spin coincidence experiments will generate
dichotomic outcomes for each single measurement.

Fig. \ref{fig3} shows the correlation function, $C\left(  \Delta p\right)  $,
for $r_{0}=4$ fm and \ for different admixtures of singlet and triplet states.
The dotted, dashed and solid lines correspond to $\mathcal{M}=$ 0.1, 0.5 and
0.9, respectively. $\mathcal{M}$ is the absolute contribution of the triplet
state. One clearly sees that different admixtures lead to very different
dependence on $\Delta p$.

\begin{figure}
[ptb]
\begin{center}
\includegraphics[
height=1.9692in,
width=3.1202in
]%
{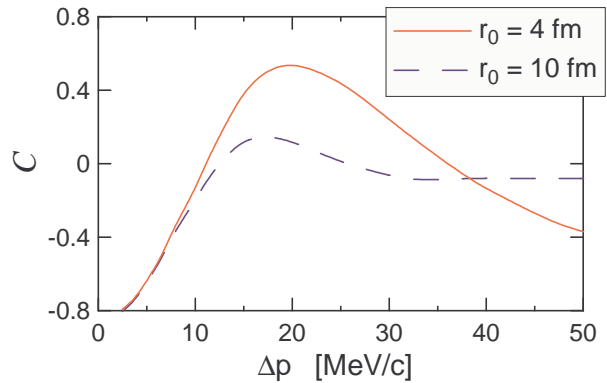}%
\caption{Correlation function, $C\left(  \Delta p\right)  $, for two-proton
triplet state decay of $^{45}$Fe with $r_{0}=4$ fm (solid curve) and
$r_{0}=10$ fm (solid curve). }%
\label{fig4}%
\end{center}
\end{figure}

Summarizing the above, we can say that the strong and Coulomb final
state interactions cannot be neglected when the volume of spatial
separation of the two-proton wavefunction is measured by $r_{0}<6$
fm. When $r_{0}<4$ fm the strong final state interaction is
noticeable in the relative motion spectrum of the two-protons (see
fig. \ref{fig1}) and its presence is reflected in the strong
reduction of the relative energy. The tidal Coulomb force due to the
charge of the daughter nucleus tends to increase considerably the
relative motion of the two protons. The same applies for the Coulomb
repulsion between the protons due to their own charge.\ These
results point to the importance of considering FSI in the
experimental analysis of two-proton decay experiments. The effect of
final state interactions are also visible in correlation functions
which are considerably modified as the initial separation of the
two-protons are probed. This is shown in fig. \ref{fig4}, where the
correlation function, $C\left(  \Delta p\right)  $, is plotted for
two-proton triplet state decay of $^{45}$Fe with $r_{0}=4$ fm (solid
curve) and $r_{0}=10$ fm (solid curve).

\begin{figure}
[ptb]
\begin{center}
\includegraphics[
height=1.9104in, width=2.9845in
]%
{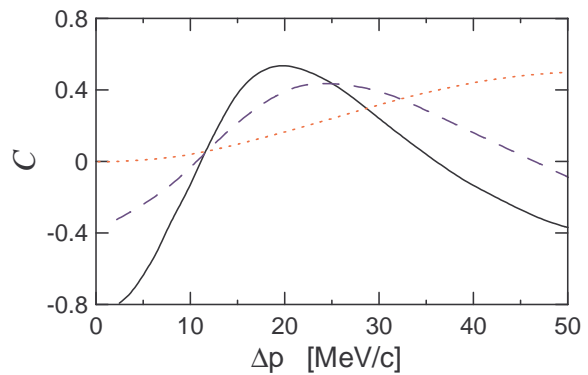}%
\caption{Correlation function, $C\left(  \Delta p\right)  $, for
$r_{0}=4$ fm and \ for different admixtures of singlet and triplet
states. The dotted, dashed and solid lines correspond to
$\mathcal{M}=$ 0.1, 0.5 and 0.9, respectively. $\mathcal{M}$ is the
absolute contribution of the triplet
state.}%
\label{fig3}%
\end{center}
\end{figure}

We have shown that correlation functions, when defined
appropriately, can clearly resolve the statistic nature of the
diproton spin state in nuclear decay. This paves another route to
study important problems of basic quantum mechanics interest, such
as the Einstein-Podolski-Rosen paradox \cite{EPR} and Scr\"odinger
cat states. Two-proton radioactivity may supply yet another test of
the EPR dilema, namely whether the spin state function provides a
complete description of quantum mechanics as we know it. These
studies would be  complementary to others performed in quantum
optics and atomic physics, as well as in some instances of nuclear
physics \cite{Lam76,Hus96}.

\begin{acknowledgments}This work was supported by the U.\thinspace S.\
Department of Energy under grant No. DE-FG02-08ER41533,
DE-FC02-07ER41457 (UNEDF, SciDAC-2) and the Brazilian agencies, CNPq
and FAPESP. M. S. Hussein is a Martin Gutzwiller Fellow 2007/2008.
\end{acknowledgments}


\begin{thebibliography}{99}


\bibitem {Bl05}B. Blank et al., Phys. Rev. Lett. 94, 2329501; ibid, 94, 249901 (2005).

\bibitem {Gi07}J. Giovinazzo et al., Phys. Rev. Lett. 99, 102501 (2007).

\bibitem {Mu07}I. Mukha et al., Phys. Rev. Lett. 99, 182501 (2007).

\bibitem {Mu06}I. Mukha et al., $Nature$, 439, 298 (2006)

\bibitem {Mi07}K. Miernik et al., Phys. Rev. Lett. 99, 192501 (2007).

\bibitem {Bai96}C.R. Bain et al., Phys. Lett. B373, 35 (1996).

\bibitem {Fyn00}H.O.U. Fynbo et al., Nucl. Phys. A677, 38 (2000).

\bibitem {Gol60}V.I. Goldansky, Nucl. Phys. 19, 482 (1960).

\bibitem {Bow91}M. G. Bowler, Phys. Lett. B270, 69 (1991).

\bibitem {BBM96}G. Baym and P. Braun-Munzinger, Nucl. Phys. A610, 286c (1996).

\bibitem {Bar98}H. W. Barz, Phys. Rev. C 53, 2536 (1996).

\bibitem {Sin98}Y. M. Sinyukov et al., Phys. Lett. B432, 248 (1998).

\bibitem {Ber03}C.A. Bertulani, J. Phys. G29, 769 (2003).

\bibitem {EPR}A. Einstein, B. Podolsky and N. Rosen, Phys. Rev. 47, 777 (1935).

\bibitem {Bell}J. Bell, Physics 1, 195 (1964).

\bibitem{Lam76} M. Lamehi-Rachti and W. Mittig,  Phys. Rev.
D 14, 2543 (1976).

\bibitem {BB94}C.A. Bertulani and G.F. Bertsch, Phys. Rev. C 49\textbf{,} 2839 (1994).

\bibitem{Hus96} M. S. Hussein,  C-Y. Lin and A. F. R. de Toledo Piza, Z. Phys. A 355, 165 (1996).

\end{thebibliography}
\end{document}